\documentclass[12pt,epsf]{article}

\usepackage{times}
\usepackage{graphicx}
\usepackage{amsfonts}
\usepackage{amsmath}
\usepackage{amssymb}
\usepackage{amstext}
\usepackage{amsthm}

\usepackage{epsfig}
\psfigdriver{dvips}
\usepackage{latexsym}
\usepackage[matrix,curve]{xypic}
\usepackage{rotating}
\rotdriver{dvips}

\begin{document}

\baselineskip 6mm
\renewcommand{\thefootnote}{\fnsymbol{footnote}}

%------------ Hyun Seok's macro's, etc  -----------

\newcommand{\nc}{\newcommand}
\newcommand{\rnc}{\renewcommand}

%\headheight=0truein
%\headsep=0truein
%\topmargin=0truein
%\oddsidemargin=0truein
%\evensidemargin=0truein
%\textheight=9truein
%\textwidth=6.5truein

\rnc{\baselinestretch}{1.24}    % 1.5 spacing btwn text lines
\setlength{\jot}{6pt}       % spacing btwn the rows of an eqnarray
\rnc{\arraystretch}{1.24}   % spacing btwn the rows of a non-eqn array

%%%%%%%%%%%%%%%%%%%%%% Equation Numbering %%%%%%%%%%%%%%%%%%%%%%%
\makeatletter
\rnc{\theequation}{\thesection.\arabic{equation}}
\@addtoreset{equation}{section}
\makeatother

%%%%%%%%%%%%%%%%%%%%%%%%%%%%%%%%%%%%%%%%%%%%%%%%%%%%%%%%%%%%%%%%%
%                                                               %
%                NEW COMMANDS AND MACROS                        %
%                                                               %
%%%%%%%%%%%%%%%%%%%%%%%%%%%%%%%%%%%%%%%%%%%%%%%%%%%%%%%%%%%%%%%%%

%%%%% Simplify some frequently used LaTeX commands %%%%%

\nc{\be}{\begin{equation}}

\nc{\ee}{\end{equation}}

\nc{\bea}{\begin{eqnarray}}

\nc{\eea}{\end{eqnarray}}

\nc{\xx}{\nonumber\\}

\nc{\ct}{\cite}

\nc{\la}{\label}

\nc{\eq}[1]{(\ref{#1})}

\nc{\newcaption}[1]{\centerline{\parbox{6in}{\caption{#1}}}}

\nc{\fig}[3]{

\begin{figure}
\centerline{\epsfxsize=#1\epsfbox{#2.eps}}
\newcaption{#3. \label{#2}}
\end{figure}
}

%%% Caligraphic letters %%%%

\def\CA{{\cal A}}
\def\CC{{\cal C}}
\def\CD{{\cal D}}
\def\CE{{\cal E}}
\def\CF{{\cal F}}
\def\CG{{\cal G}}
\def\CH{{\cal H}}
\def\CK{{\cal K}}
\def\CL{{\cal L}}
\def\CM{{\cal M}}
\def\CN{{\cal N}}
\def\CO{{\cal O}}
\def\CP{{\cal P}}
\def\CR{{\cal R}}
\def\CS{{\cal S}}
\def\CU{{\cal U}}
\def\CV{{\cal V}}
\def\CW{{\cal W}}
\def\CY{{\cal Y}}
\def\CZ{{\cal Z}}

%%% Double line letters %%%

\def\IB{{\hbox{{\rm I}\kern-.2em\hbox{\rm B}}}}
\def\IC{\,\,{\hbox{{\rm I}\kern-.50em\hbox{\bf C}}}}
\def\ID{{\hbox{{\rm I}\kern-.2em\hbox{\rm D}}}}
\def\IF{{\hbox{{\rm I}\kern-.2em\hbox{\rm F}}}}
\def\IH{{\hbox{{\rm I}\kern-.2em\hbox{\rm H}}}}
\def\IN{{\hbox{{\rm I}\kern-.2em\hbox{\rm N}}}}
\def\IP{{\hbox{{\rm I}\kern-.2em\hbox{\rm P}}}}
\def\IR{{\hbox{{\rm I}\kern-.2em\hbox{\rm R}}}}
\def\IZ{{\hbox{{\rm Z}\kern-.4em\hbox{\rm Z}}}}

%%% Greek letters %%%

\def\a{\alpha}
\def\b{\beta}
\def\d{\delta}
\def\ep{\epsilon}
\def\ga{\gamma}
\def\k{\kappa}
\def\l{\lambda}
\def\s{\sigma}
\def\t{\theta}
\def\w{\omega}
\def\G{\Gamma}

%%%%% Mathematical Symbols

\def\half{\frac{1}{2}}
\def\dint#1#2{\int\limits_{#1}^{#2}}
\def\goto{\rightarrow}
\def\para{\parallel}
\def\brac#1{\langle #1 \rangle}
\def\curl{\nabla\times}
\def\div{\nabla\cdot}
\def\p{\partial}

%%%%% Roman pont in math

\def\Tr{{\rm Tr}\,}
\def\det{{\rm det}}

%%%%% Special Letters

\def\vare{\varepsilon}
\def\zbar{\bar{z}}
\def\wbar{\bar{w}}
\def\what#1{\widehat{#1}}

%%%%% For this paper only

\def\ad{\dot{a}}
\def\bd{\dot{b}}
\def\cd{\dot{c}}
\def\dd{\dot{d}}
\def\so{SO(4)}
\def\bfr{{\bf R}}
\def\bfc{{\bf C}}
\def\bfz{{\bf Z}}

\begin{titlepage}

%---------------- preprint number ---------------

\hfill\parbox{3.7cm} {HU-EP-07/12 \\
{\tt arXiv:0704.0929}}

\vspace{15mm}

\begin{center}
%------------------------ title ------------------------
{\Large \bf  Noncommutative Electromagnetism As A Large $N$ Gauge Theory}

\vspace{10mm}
%---------------- authors and addresses ----------------
Hyun Seok Yang \footnote{hsyang@physik.hu-berlin.de}
\\[10mm]

{\sl Institut f\"ur Physik, Humboldt Universit\"at zu Berlin \\
Newtonstra\ss e 15, D-12489 Berlin, Germany}

\end{center}

\thispagestyle{empty}

\vskip1cm

%----------------------- abstract ----------------------

\centerline{\bf ABSTRACT}
\vskip 4mm
\noindent
We map noncommutative (NC) $U(1)$ gauge theory on $\IR^d_C \times
\IR^{2n}_{NC}$ to $U(N \to \infty)$ Yang-Mills theory on $\IR^d_C$, where
$\IR^d_C$ is a $d$-dimensional commutative spacetime while $\IR^{2n}_{NC}$ is a
$2n$-dimensional NC space. The resulting $U(N)$ Yang-Mills
theory on $\IR^d_C$ is equivalent to that obtained by the dimensional
reduction of $(d+2n)$-dimensional $U(N)$ Yang-Mills theory onto
$\IR^d_C$. We show that the gauge-Higgs system $(A_\mu,\Phi^a)$ in the
$U(N \to \infty)$ Yang-Mills theory on $\IR^d_C$ leads to an emergent
geometry in the $(d+2n)$-dimensional spacetime whose metric was determined by
Ward a long time ago. In particular, the 10-dimensional gravity for $d=4$ and
$n=3$ corresponds to the emergent geometry arising from the 4-dimensional
$\CN =4$ vector multiplet in the AdS/CFT duality. We further elucidate the emergent
gravity by showing that the gauge-Higgs system $(A_\mu,\Phi^a)$ in half-BPS
configurations describes self-dual Einstein gravity.
\\



\vspace{1cm}

\today

\end{titlepage}

\renewcommand{\thefootnote}{\arabic{footnote}}
\setcounter{footnote}{0}

\section{Introduction}

A noncommutative (NC) spacetime $M$ is obtained by introducing a symplectic
structure $B = \half B_{ab} dy^a \wedge dy^b$ and then by quantizing
the spacetime with its Poisson structure $\theta^{ab} \equiv (B^{-1})^{ab}$,
treating it as a quantum phase space. That is, for $f,g \in C^\infty(M)$,
\be \la{poisson-bracket}
\{f, g\} = \theta^{ab} \left(\frac{\p f}{\p y^a} \frac{\p g}{\p y^b}
- \frac{\p f}{\p y^b} \frac{\p g}{\p y^a}\right) \Rightarrow
- i[\what{f},\what{g}].
\ee
According to the Weyl-Moyal map \ct{nc-review,szabo}, the NC algebra of operators
is equivalent to the deformed algebra of functions defined
by the Moyal $\star$-product, i.e.,
\begin{equation}\label{star-product}
\what{f} \cdot \what{g} \cong (f \star g)(y) = \left.\exp\left(\frac{i}{2}
\theta^{ab} \partial_{a}^{y}\partial_{b}^{z}\right)f(y)g(z)\right|_{y=z}.
\end{equation}
Through the quantization rules \eq{poisson-bracket} and \eq{star-product},
one can define NC $\IR^{2n}$ by the following commutation relation
\be \la{nc-spacetime}
[y^a, y^b]_\star = i \theta^{ab}.
\ee

It is well-known \ct{szabo,mpla} that a NC field theory can be identified basically
with a matrix model or a large $N$ field theory.
This claim is based on the following fact.
Let us consider a NC $\IR^2$ for simplicity,
\begin{equation} \label{nc-plane}
    [x,y] = i \theta,
\end{equation}
although the same argument equally holds for a NC $\IR^{2n}$ as it
will be shown later. After scaling the coordinates $ x \to
\sqrt{\theta} x, \; y \to \sqrt{\theta} y$, the NC plane
\eq{nc-plane} becomes the Heisenberg algebra of harmonic
oscillator
\begin{equation} \label{oscillator}
    [a, a^\dagger] = 1.
\end{equation}
It is a well-known fact from quantum mechanics that
the representation space of NC $\IR^2$ is given by an infinite-dimensional,
separable Hilbert space $\CH = \{ |n \rangle, \; n=0,1,
\cdots \}$ which is orthonormal, i.e., $\langle n| m \rangle = \delta_{nm}$
and complete, i.e.,  $ \sum_{n=0}^\infty |n \rangle \langle n| = 1$.
Therefore a scalar field $\widehat{\phi} \in \CA_\theta $ on the
NC plane \eq{nc-plane} can be expanded in terms of the complete operator basis
\begin{equation}\label{matrix-basis}
\CA_\theta = \{ |m \rangle \langle n|, \; n,m = 0,1, \cdots \},
\end{equation}
that is,
\begin{equation}\label{op-matrix}
    \widehat{\phi}(x,y) = \sum_{n,m} M_{mn} |m \rangle \langle n|.
\end{equation}
One can regard $M_{mn}$ in \eq{op-matrix} as components of an $N \times N$
matrix $M$ in the $N \to \infty$ limit.
More generally one may replace NC $\IR^2$ by a Riemann surface $\Sigma_g$
of genus $g$ which can be quantized via deformation quantization \ct{kontsevich}.
For a compact Riemann surface $\Sigma_g$ with finite area
$A(\Sigma_g)$, the matrix representation can be finite-dimensional, e.g., for
a fuzzy sphere \ct{fuzzy-sphere}.
In this case, $A(\Sigma_g) \sim \theta N$ but we simply take
the limit $ N \to \infty$.
We then arrive at the well-known relation:
\be \la{sun-sdiff}
\mathrm{Scalar \; field \; on \; NC} \; \IR^2 \;
(\mathrm{or} \; \Sigma_g) \Longleftrightarrow  N \times N \;
\mathrm{matrix} \; \mathrm{at} \; N \to \infty.
\ee
If $\widehat{\phi}$ is a real scalar field, then $M$ should be a
Hermitean matrix. We will see that the above relation \eq{sun-sdiff}
has far-reaching applications to string theory.

The matrix representation \eq{op-matrix} clarifies why NC $U(1)$
gauge theory is a large $N$ gauge theory. An important point is that
the NC gauge symmetry acts as a unitary transformation on $\CH$ for
a field $\widehat{\phi} \in \CA_\theta $ in the adjoint
representation of $U(1)$ gauge group
\be \la{ad-scalar}
\widehat{\phi} \to U  \widehat{\phi} \, U^\dagger.
\ee
This NC gauge symmetry $U_{\rm{cpt}}(\mathcal{H})$
is so large that $U_{\rm{cpt}}({\mathcal{H}}) \supset U(N)
\;(N \rightarrow \infty)$ \cite{nc-group,nc-top}, which is rather obvious in the
matrix basis \eq{matrix-basis}. Therefore the NC gauge theory is
essentially a large $N$ gauge theory. It becomes more precise on a
NC torus through the Morita equivalence where NC $U(1)$ gauge theory
with rational $\theta = M/N$ is equivalent to an ordinary $U(N)$
gauge theory \ct{sw}. For this reason, it is not so surprising that
NC electromagnetism shares essential properties appearing in a large
$N$ gauge theory such as $SU(N \to \infty)$ Yang-Mills theory or
matrix models.

It is well-known \ct{1/N} that $1/N$ expansion
of any large $N$ gauge theory using the double line formalism reveals
a picture of a topological expansion in terms of surfaces of different
genus, which can be interpreted in terms of closed string variables
as the genus expansion of string amplitudes. It has been underlain
the idea that large $N$ gauge theories have a dual description in
terms of gravitational theories in higher dimensions. For example,
BFSS matrix model \ct{bfss}, IKKT matrix model \ct{ikkt}
and AdS/CFT duality \ct{ads-cft}.
From the perspective \eq{sun-sdiff}, the $1/N$ expansion corresponds
to the NC deformation in terms of $\theta/A(\Sigma_g)$.

All these arguments imply that there exists a solid map between a NC
gauge theory and a large $N$ gauge theory. In this work we will find
a sound realization of this idea. It turns out that the emergent
gravity recently found in \ct{sty,ys,hsy1,hsy2} can be elegantly
understood in this framework. Therefore the correspondence between
NC field theory and gravity \ct{mpla} is certainly akin to the
gauge/gravity duality in large $N$ limit \ct{bfss,ikkt,ads-cft}.

This paper is organized as follows. In Section 2 we map NC $U(1)$
gauge theory on $\IR^d_C \times \IR^{2n}_{NC}$ to $U(N \to \infty)$
Yang-Mills theory on $\IR^d_C$, where $\IR^d_C$ is a $d$-dimensional
commutative spacetime while $\IR^{2n}_{NC}$ is a $2n$-dimensional NC
space. The resulting $U(N)$ Yang-Mills theory on $\IR^d_C$ is
equivalent to that obtained by the dimensional reduction of
$(d+2n)$-dimensional $U(N)$ Yang-Mills theory onto $\IR^d_C$. In
Section 3, we show that the gauge-Higgs system $(A_\mu,\Phi^a)$ in
the $U(N \to \infty)$ Yang-Mills theory on $\IR^d_C$ leads to an
emergent geometry in the $(d+2n)$-dimensional spacetime whose metric
was determined by Ward \ct{ward} a long time ago. In particular, the
10-dimensional gravity for $d=4$ and $n=3$ corresponds to the
emergent geometry arising from the 4-dimensional $\CN =4$ vector
multiplet in the AdS/CFT duality \ct{ads-cft}. We further elucidate
the emergent gravity in Section 4 by showing that the gauge-Higgs
system $(A_\mu,\Phi^a)$ in half-BPS configurations describes
self-dual Einstein gravity. A notable point is that the emergent
geometry arising from the gauge-Higgs system $(A_\mu,\Phi^a)$ is
closely related to the bubbling geometry in AdS space found in
\ct{llm}. Finally, in Section 5, we discuss several interesting
issues that naturally arise from our construction.

\section{A Large $N$ Gauge Theory From NC $U(1)$ Gauge Theory}

We will consider a NC $U(1)$ gauge theory on $\IR^D = \IR^d_C \times
\IR^{2n}_{NC}$, where $D$-dimensional coordinates $X^M \; (M=1,\cdots, D)$ are
decomposed into $d$-dimensional commutative ones, denoted as $z^\mu \;
(\mu=1, \cdots, d)$ and $2n$-dimensional NC ones, denoted as $y^a \;
(a = 1, \cdots, 2n)$, satisfying the relation \eq{nc-spacetime}.
We assume the metric on $\IR^D = \IR^d_C \times \IR^{2n}_{NC}$ as the
following form \footnote{\label{metric-convention}
Here we can take the $d$-dimensional spacetime
metric $g_{\mu\nu}$ with either Lorentzian or Euclidean
signature since the signature is inconsequential in our most
discussions. But we implicitly assume the Euclidean signature for
some other discussions.}
\bea \la{flat-metric}
ds^2 &=& \CG_{MN} dX^M dX^N \xx
&=& g_{\mu\nu} dz^\mu dz^\nu + G_{ab} dy^a dy^b.
\eea
The action for $D$-dimensional NC $U(1)$ gauge theory is given by
\be \la{starting-action}
S = \frac{1}{4 g^2_{YM}} \int d^D X \sqrt{\det \CG} \CG^{MP} \CG^{NQ}
(F_{MN} + \Phi_{MN}) \star (F_{PQ} + \Phi_{PQ}),
\ee
where the NC field strength $F_{MN}$ is defined by
\be \la{nc-field}
F_{MN} = \p_M A_N - \p_N A_M -i [ A_M, A_N]_\star.
\ee
The constant two-form $\Phi$ will be taken either $0$ or $-B = - \half
B_{ab} dy^a \wedge dy^b$ with ${\rm rank} (B) = 2n$.

Here we will use the background independent prescription \ct{sw,seiberg} where
the open string metric $G_{ab}$, the noncommutativity $\theta^{ab}$ and the
open string coupling $G_s$ are determined by
\be \la{open}
\theta^{ab} = \Bigl(\frac{1}{B}\Bigr)^{ab}, \quad
G_{ab} = - \kappa^2 \Bigl( B \frac{1}{g} B \Bigr)_{ab}, \quad
G_s = g_s \sqrt{\det^\prime (\kappa B g^{-1})},
\ee
with $\kappa \equiv 2 \pi \alpha^\prime$. The closed string metric $g_{ab}$ in
Eq.\eq{open} is independent of $g_{\mu\nu}$ in Eq.\eq{flat-metric} and
$\det^\prime$ denotes a determinant taken along NC directions
only in $\IR^{2n}_{NC}$. In terms of these parameters, the couplings are
related by
\bea \la{coupling1}
&& \frac{1}{g_{YM}^2} = \frac{\kappa^{\frac{4-D}{2}}}{(2\pi)^{\frac{D-2}{2}}
  G_s}, \\
&& \frac{\sqrt{\det^\prime G}}{G_s} = \frac{\kappa^n}{g_s |{\rm Pf} \theta|}.
\eea

An important fact is that translations in NC directions are basically gauge transformations, i.e.,
$e^{ik \cdot y} \star f(z,y) \star e^{-ik \cdot y} = f(z, y +  \theta \cdot k)$
for any $f(z,y) \in C^\infty(M)$. This means that translations along NC
directions act as inner derivations of the NC algebra $\CA_\theta$:
\be \la{inner-der}
[y^a, f]_\star = i \theta^{ab} \p_b f.
\ee
Using this relation, each component of $F_{MN}$ can be written as the
following forms
\bea \la{fcc}
&& F_{\mu\nu} = i [D_\mu, D_\mu]_\star, \\
\la{fcnc}
&& F_{\mu a} = \theta^{-1}_{ab} [D_\mu, x^b]_\star = -F_{a \mu}, \\
\la{fncnc}
&& F_{a b} = - i \theta^{-1}_{ac} \theta^{-1}_{bd} \Bigl([x^c, x^d]_\star - i
\theta^{cd} \Bigr),
\eea
where the covariant derivative $D_\mu$ and the covariant coordinate
$x^a$ are, respectively, defined by
\bea \la{cov-der}
&& D_\mu \equiv \p_\mu - i A_\mu, \\
\la{cov-x}
&& x^a \equiv y^a + \theta^{ab} A_b.
\eea

Collecting all these facts, one gets the following expression for
the action \eq{starting-action} with $\Phi= -B$ \footnote{If $\Phi
=0$ in Eq.\eq{starting-action}, the only change in
Eq.\eq{matrix-action} is $[\Phi^a, \Phi^b] \to [\Phi^a, \Phi^b] -
\frac{i}{\kappa^2}\theta^{ab}$.}
\bea \la{matrix-action}
S &=& \frac{(2\pi\kappa)^{\frac{4-d}{2}}}{2\pi g_s} \int d^d z
\sqrt{\det g_{\mu\nu}} \Tr_{\CH} \left( \frac{1}{4} g^{\mu\lambda}
  g^{\nu\sigma} F_{\mu\nu} \star F_{\lambda\sigma} + \frac{1}{2} g^{\mu\nu}
g_{ab} D_\mu \Phi^a \star D_\nu \Phi^b \right. \xx
&& \hspace{5cm} \left. - \frac{1}{4}g_{ac}g_{bd}[\Phi^a, \Phi^b]_\star
\star [\Phi^c, \Phi^d]_\star \right),
\eea
where we defined adjoint scalar fields $\Phi^a \equiv x^a/ \kappa$ of mass
dimension and
\be \la{trace}
 \Tr_{\CH} \equiv \int \frac{d^{2n}y}{(2\pi)^n |{\rm Pf} \theta|}.
\ee
Note that the number of the adjoint scalar fields is equal to the rank of
$\theta^{ab}$. The resulting action \eq{matrix-action} is not new but
rather well-known in NC field theory, e.g., see \ct{seiberg,japan}.

The NC algebra \eq{nc-spacetime} is equivalent to the Heisenberg algebra of
an $n$-dimensional harmonic oscillator in a frame where $\theta^{ab}$ has a
canonical form:
\be \la{nc-harmonic}
[a_i, a_j^\dagger] = \delta_{ij}, \quad (i,j=1, \cdots, n).
\ee
The NC space \eq{nc-spacetime} is therefore represented by the
infinite-dimensional Hilbert space $\CH = \{ |\vec{m} \rangle \equiv
|m_1, \cdots, m_n \rangle; m_i = 0,1,\cdots,N \to \infty \; {\rm
for} \; i=1, \cdots, n \}$ whose set of eigenvalues forms an
$n$-dimensional positive integer lattice. A set of operators in
$\CH$
\be \la{op-basis}
\CA_\theta = \{ |\vec{m} \rangle \langle \vec{n} |; m_i, n_i = 0,1,\cdots,N \to \infty \;
{\rm for} \; i=1, \cdots, n \}
\ee
can be identified with the generators of a complete operator basis
and so any NC field $\phi(z,y) \in \CA_\theta$ can be expanded in
the basis
\eq{op-basis} as follows,
\be \la{op-expansion}
\phi(z,y) = \sum_{\vec{m},\vec{n}} \Omega_{\vec{m},\vec{n}} \; (z)
|\vec{m} \rangle \langle \vec{n} |.
\ee

Now we use the `Cantor diagonal method' to put the $n$-dimensional positive
integer lattice in $\CH$ into a one-to-one correspondence
with the infinite set of natural numbers (i.e., $1$-dimensional positive
integer lattice): $|\vec{m} \rangle \leftrightarrow |i\rangle, \; i=1,\cdots,N
\to \infty $. In this one-dimensional basis, Eq.\eq{op-expansion} is
relabeled as the following form
\be \la{matrix-expansion}
\phi(z,y) = \sum_{i,j} \Omega_{ij} \; (z)
|i \rangle \langle j |.
\ee
Following the motivation discussed in the Introduction, we regard
$\Omega_{ij}(z)$ in \eq{matrix-expansion} as components of an $N
\times N$ matrix $\Omega$ in the $N \to \infty$ limit, which also
depend on $z^\mu$, the coordinates of $\IR_C^d$. If the field
$\phi(z,y)$ is real which is the case for the gauge-Higgs system
$(A_\mu,\Phi^a)$ in the action \eq{matrix-action}, the matrix
$\Omega$ should be Hermitean, but not necessarily traceless. So the
$N \times N$ matrix $\Omega(z)$ can be regarded as a field in $U(N
\to \infty)$ gauge theory on $d$-dimensional commutative space
$\IR^d_C$, where $\Tr_{\CH}$ in \eq{trace} is identified with the
matrix trace over the basis \eq{matrix-expansion}. All the
dependence on NC coordinates is now encoded into $N \times N$
matrices and the noncommutativity in terms of star product is
transferred to the matrix product.

Adopting the matrix representation \eq{matrix-expansion}, the
$D$-dimensional NC $U(1)$ gauge theory
\eq{starting-action} is mapped to the $U(N \to \infty)$ Yang-Mills theory
on $d$-dimensional commutative space $\IR^d_C$. One can see that the
resulting $U(N)$ Yang-Mills theory on $\IR^d_C$ in
Eq.\eq{matrix-action} is equivalent to that obtained by the
dimensional reduction of $(d+2n)$-dimensional $U(N)$ Yang-Mills
theory onto $\IR^d_C$. It might be emphasized that the map between
the $D$-dimensional NC $U(1)$ gauge theory and the $d$-dimensional
$U(N \to \infty)$ Yang-Mills theory is ``exact" and thus the two
theories should describe a completely equivalent physics. For
example, we can recover the $D$-dimensional NC $U(1)$ gauge theory
on $\IR^d_C \times \IR^{2n}_{NC}$ from the $d$-dimensional $U(N \to
\infty)$ Yang-Mills theory on $\IR^d_C$ by recalling that the number
of adjoint Higgs fields in the $U(N)$ Yang-Mills theory is equal to
the dimension of the extra NC space $\IR^{2n}_{NC}$ and by applying
the dictionary in Eqs.\eq{fcc}-\eq{fncnc}.

One can introduce linear algebraic conditions of $D$-dimensional
field strengths $F_{MN}$ as a higher dimensional analogue of
$4$-dimensional self-duality equations such that the Yang-Mills
equations in the action \eq{starting-action} follow automatically.
These are of the following type \ct{high-sd,high-ward}
\be \la{high-sd}
\half T_{MNPQ} F_{PQ} = \lambda F_{MN}
\ee
with a constant 4-form tensor $T_{MNPQ}$. The relation \eq{high-sd} clearly
implies via the Bianchi identity $D_{[M} F_{PQ]}=0$ that
the Yang-Mills equations are satisfied provided $\lambda$ is nonzero.
For $D > 4$, the 4-form tensor $T_{MNPQ}$ cannot be invariant under $SO(D)$
transformations and the equation \eq{high-sd} breaks the rotational symmetry
to a subgroup $H \subset SO(D)$. Thus the resulting first order equations
can be classified by the unbroken symmetry $H$ under which $T_{MNPQ}$ remain
invariant \ct{high-sd,high-ward}.
It was also shown \ct{bkp} that the first order linear equations above are
closely related to supersymmetric states, i.e., BPS states in
higher dimensional Yang-Mills theories.

The equivalence between $D$- and $d$-dimensional gauge theories can
be effectively used to classify classical solutions in the
$d$-dimensional $U(N)$ Yang-Mills theory \eq{matrix-action}. The
group theoretical classification \ct{high-sd}, integrability
condition \ct{high-ward} and BPS states \ct{bkp} for the
$D$-dimensional first-order equations \eq{high-sd} can be directly
translated into the properties of the gauge-Higgs system $(A_\mu,
\Phi^a)$ in the $d$-dimensional $U(N)$ gauge theory
\eq{matrix-action}. These classifications will also be useful to
classify the geometries emerging from the gauge-Higgs system
$(A_\mu,\Phi^a)$ in the $U(N \to \infty)$ Yang-Mills theory
\eq{matrix-action}, which will be discussed in the next section.
Unfortunately, the $D=10$ case is missing in \ct{high-sd,high-ward,bkp}
which is the most interesting case ($d=4$ and $n=3$) related to the AdS/CFT duality.

\section{Emergent Geometry From NC Gauge Theory}

Let us first recapitulate the result in \ct{ward}. It turns out that
the Ward's construction perfectly fits with the emergent geometry
arising from the gauge-Higgs system $(A_\mu,\Phi^a)$ in the $U(N \to
\infty)$ Yang-Mills theory \eq{matrix-action}. Suppose that we have
gauge fields on $\IR^d_C$ taking values in the Lie algebra of
volume-preserving vector fields on an $m$-dimensional manifold $M$
\ct{ashtekar,mason-newman}. In other words, the gauge group $G =
SDiff(M)$. The gauge covariant derivative is given by
Eq.\eq{cov-der}, but the $A_\mu(z)$ are now vector fields on $M$,
also depending on $z^\mu \in \IR^d_C$. The other ingredient in
\ct{ward} consists of $m$ Higgs fields $\Phi^a(z) \in sdiff(M)$, the
Lie algebra of $SDiff(M)$,  for $a=1,\cdots,m$. The idea
\ct{ashtekar,mason-newman} is to specify that
\be \la{on-frame}
f^{-1}(D_1, \cdots, D_d, \Phi_1, \cdots, \Phi_m)
\ee
forms an orthonormal frame and hence defines a metric on $\IR^d_C
\times M$ with a volume form $\nu = d^d z \wedge \omega$. Here $f$ is
a scalar, a conformal factor, defined by
\be \la{v-form}
f^2 = \omega (\Phi_1, \cdots, \Phi_m).
\ee

The result in \ct{ashtekar,mason-newman} immediately implies that
the gauge-Higgs system $(A_\mu,\Phi^a)$ leads to a metric on the
$(d+m)$-dimensional space $\IR^d_C \times M$. A local coordinate
expression for this metric is easily obtained from Eq.\eq{on-frame}.
Let $y^a$ be local coordinates on $M$. So $A_\mu(z)$ and $\Phi_a(z)$
have the form
\be \la{local-vector}
A_\mu(z) = A_\mu^a(z,y) \frac{\p}{\p y^a}, \qquad
\Phi_a(z) = \Phi_a^b(z,y) \frac{\p}{\p y^b},
\ee
where the $y$-dependence, originally hidden in the Lie algebra of $G
= SDiff(M)$, now explicitly appears in the coefficients $A_\mu^a$
and $\Phi_a^b$. Let $V^a_b$ denote the inverse of the $m \times m$
matrix $\Phi^b_a$, and let ${\bf A}^a$ denote the 1-form $A^a_\mu
dz^\mu$. Then the metric is \ct{ward}
\be \la{emergent-metric}
ds^2 = f^2 \delta_{\mu\nu} dz^\mu dz^\nu + f^2 \delta_{ab}
V^a_c V^b_d (dy^c - {\bf A}^c) (dy^d - {\bf A}^d).
\ee
It will be shown later that the choice of the volume form $\omega$
for the conformal factor \eq{v-form} corresponds to that of
a particular conformally flat background although we mostly assume
a flat volume form, i.e., $\omega \sim  dy^1 \wedge \cdots \wedge dy^{2n}$,
unless explicitly specified.

The gauge and Higgs fields in Eq.\eq{local-vector} are not arbitrary
but must be subject to the Yang-Mills equations, for example,
derived from the action \eq{matrix-action}, which are, in most
cases, not completely integrable. Hence to completely determine the
geometric structure emerging from the gauge-Higgs system
$(A_\mu,\Phi^a)$ is as much difficult as solving the Einstein
equations in general. But the self-dual Yang-Mills equations in four
dimensions or Eq.\eq{high-sd} in general are, in some sense,
``completely solvable''. Thus the metric \eq{emergent-metric} for
these cases might be completely determined. Let us discuss two
notable examples. See \ct{ward} for more examples describing
4-dimensional self-dual Einstein gravity.

$\bullet$ {\it Case} $d=0, \; m=4$: This case was dealt with in detail in
\ct{mason-newman,cmn,joyce}. It was proved that the self-dual Einstein
equations are equivalent to the self-duality equations
\be \la{sd-higgs}
[\Phi_a, \Phi_b] = \pm \half \vare_{abcd} [\Phi_c, \Phi_d]
\ee
on the four Higgs fields $\Phi_a$. Furthermore reinterpreting $n$ of
the $\Phi_a$'s as $D_\mu$ leads to the case $d=n, \; m=4-n$. In
Section 5, we will discuss the physical meaning about the
interpretation $\Phi_a \mapsto D_\mu$.

$\bullet$ {\it Case} $d=3, \; m=1$: Here $M$ is one-dimensional, so
the Lie algebra of vector fields on $M$ is the Virasoro algebra.
Thus $A_\mu$ and $\Phi$ are now real-valued vector fields on $M$
which must be independent of $y$ to preserve the volume form $\nu =
d^3 z \wedge dy$ \ct{joyce}. The metric \eq{emergent-metric} reduces
to
\be \la{gibbons-hawking}
ds^2 = \Phi d\vec{z} \cdot d\vec{z} + \Phi^{-1} (dy- A_\mu dz^\mu)^2
\ee
and has a Killing vector $\p/\p y$.
In this case, the self-duality equations \eq{sd-higgs} reduce to
the Abelian Bogomol'nyi equations, $\nabla \times \vec{A} = \nabla \Phi$,
and the metric \eq{gibbons-hawking} describes a gravitational instanton
\ct{gibb-hawk}.

Recently we showed in \ct{hsy1,hsy2} for the $d=0$ and $m=4$ case
that self-dual electromagnetism in NC spacetime is equivalent to
self-dual Einstein gravity and the metric is precisely given by
Eq.\eq{emergent-metric}. A key observation \ct{hsy2} was that the
self-dual system \eq{sd-higgs} defined by vector fields on $M$ can
be derived from the action
\eq{starting-action} or \eq{matrix-action} for slowly varying
fields, where all $\star$-commutators between NC fields are
approximated by the Poisson bracket \eq{poisson-bracket}. An
important point in NC geometry is that the adjoint action of
(covariant) coordinates with respect to star product can be
identified with (generalized) vector fields on some (curved)
manifold \ct{hsy1,hsy2}, as the trivial case was already used in
Eq.\eq{inner-der}. In the end, a D-dimensional manifold described by
the metric \eq{emergent-metric} corresponds to an emergent geometry
arising from the gauge-Higgs system in Eq.\eq{local-vector}. Now we
will show in a general context how the nontrivial geometry
\eq{emergent-metric} emerges from the gauge-Higgs system
$(A_\mu,\Phi^a)$ in the action
\eq{matrix-action}.

Let us collectively denote the covariant derivatives $D_\mu$ in
\eq{cov-der} and the Higgs fields $D_a \equiv -i \kappa B_{ab} \Phi^b = -i
(B_{ab} y^b + A_a)$ in \eq{cov-x} as $D_A(z,y)$. Therefore
$D_A(z,y)$ transform covariantly under NC $U(1)$ gauge
transformations
\be \la{gauge-tr}
D_A(z,y) \to g(z,y)\star D_A(z,y) \star g^{-1}(z,y).
\ee
Define the adjoint action of $D_A(z,y)$ with respect to star product
acting on any NC field $f(z,y) \in  \CA_\theta$:
\be \la{adjoint-action}
{\rm ad}_{D_A} [f] \equiv [D_A, f]_\star.
\ee
Then it is easy to see \ct{hsy2} that the above adjoint action satisfies the
Leibniz rule and the Jocobi identity, i.e.,
\bea \la{leibniz}
&& [D_A, f \star g]_\star = [D_A, f]_\star \star g +
f \star [D_A, g]_\star,\\
\la{jacobi}
&&  [D_A,[D_B, f]_\star ]_\star -  [D_B,[D_A, f]_\star ]_\star =
[[D_A, D_B]_\star, f]_\star.
\eea
These properties imply that ${\rm ad}_{D_A}$ can be identified with
`generalized' vector fields or Lie derivatives acting on the algebra
$\CA_\theta$, which can be viewed as a gauge covariant
generalization of the inner derivation \eq{inner-der}. Note that the
generalized vector field in Eq.\eq{adjoint-action} is a kind of
general higher order differential operators in \ct{nc-gravity}.
Indeed it turns out that they constitute a generalization of volume
preserving diffeomorphisms to $\star$-differential operators acting
on $\CA_\theta$ (see Eqs.(4.1) and (4.2) in \ct{nc-top}).

In particular, the generalized vector fields in
Eq.\eq{adjoint-action} reduce to usual vector fields in the
commutative, i.e. $\CO(\theta)$, limit:
\bea \la{semi-vector}
{\rm ad}_{D_A} [f] & = & i \theta^{ab} \frac{\p D_A}{\p y^a}
\frac{\p f}{\p y^b} + \cdots = i \{ D_A, f \} + \CO(\theta^3) \xx
&\equiv& V_A^a(z,y) \p_a f(z,y) + \CO(\theta^3)
\eea
where we defined $[\p_\mu, f]_\star = \p_\mu f$. Note that the
vector fields $V_A(z,y)= V_A^a(z,y) \p_a $ are exactly of the same
form as Eq.\eq{local-vector} and belong to the Lie algebra of volume
preserving diffeomorphisms, as precisely required in the Ward
construction \eq{on-frame}, since they are all divergence free,
i.e., $\p_a V_A^a = 0$. Thus the vector fields $ f^{-1} V_A(z,y)$
for $A = 1, \cdots, D$ can be identified with the orthonormal frame
\eq{on-frame} defining the metric \eq{emergent-metric}. It should be
emphasized that the emergent gravity \eq{emergent-metric} arises
from a general, not necessarily self-dual, gauge-Higgs system
$(A_\mu,\Phi^a)$ in the action \eq{matrix-action}.

Note that
\be \la{comm-f}
[D_A, D_B]_\star = -i (F_{AB} - B_{AB})
\ee
where the NC field strength $F_{AB}$ is given by Eq.\eq{nc-field}.
Then the Jacobi identity \eq{jacobi} leads to the following identity
for a constant $B_{AB}$
\be \la{comm-joc}
{\rm ad}_{[D_A, D_B]_\star} = -i \; {\rm ad}_{F_{AB}} = [{\rm
ad}_{D_A}, {\rm ad}_{D_B}]_\star .
\ee
The inner derivation \eq{semi-vector} in commutative limit is
reduced to the well-known map $C^\infty(M) \to TM: f \mapsto X_f$
between the Poisson algebra $(C^\infty(M), \{ \cdot, \cdot \})$ and
vector fields in $TM$ defined by $X_f(g) = \{ g, f \}$ for any
smooth function $g \in C^\infty(M)$. The Jacobi identity for the
Poisson algebra $(C^\infty(M), \{ \cdot, \cdot \})$ then leads to
the Lie algebra homomorphism
\be \la{lie-homo}
X_{\{ f, g \}} = - [X_f, X_g]
\ee
where the right-hand side is defined by the Lie bracket between
Hamiltonian vector fields. One can check by identifying $f = D_A$
and $g = D_B$ that the Lie algebra homomorphism \eq{lie-homo}
correspond to the commutative limit of the Jacobi identity
\eq{jacobi}. That is, one can deduce from Eq.\eq{lie-homo} the
following identity
\be \la{fvec-homo}
X_{F_{AB}} = - [V_A, V_B]
\ee
using the relation $\{ D_A, D_B \} = - F_{AB} + B_{AB}$ and $X_{
D_A} = i V_A$.

Using the homomorphism \eq{fvec-homo}, one can translate the
generalized self-duality equation \eq{high-sd} into the structure
equation between vector fields
\be \la{general-sd}
\half T_{ABCD} F_{CD} = \lambda F_{AB} \quad \Leftrightarrow \quad
\half T_{ABCD} [V_C, V_D] = \lambda  [V_A, V_B].
\ee
Therefore a D-dimensional NC gauge field configuration satisfying
the first-order system defined by the left-hand side of
Eq.\eq{general-sd} is isomorphic to a D-dimensional emergent
geometry defined by the right-hand side of Eq.\eq{general-sd} whose
metric is given by Eq.\eq{emergent-metric}. For example, in four
dimensions where $T_{ABCD} = \varepsilon_{ABCD}$ and $\lambda = \pm
1$, the right-hand side of Eq.\eq{general-sd} is precisely equal to
Eq.\eq{sd-higgs} describing gravitational instantons
\ct{ashtekar,mason-newman,cmn,joyce}. This proves, as first shown in
\ct{hsy1,hsy2}, that self-dual NC electromagnetism is equivalent to
self-dual Einstein gravity. Note that the Einstein gravity described
by the metric \eq{emergent-metric} arises from the commutative,
i.e., $\CO(\theta)$ limit. Therefore it is natural to expect that
the higher order differential operators in Eq.\eq{semi-vector}, e.g.
$\CO(\theta^3)$, give rise to higher order gravity \ct{hsy2}. We
will further discuss the derivative correction in Section 5.

The 10-dimensional metric \eq{emergent-metric} for $d=4$ and $n=3
\;(m=6)$ is particularly interesting since it corresponds to an
emergent geometry arising from the 4-dimensional $\CN =4$ vector
multiplet in the AdS/CFT duality. Note that the gravity in the
AdS/CFT duality is an emergent phenomenon arising from particle
interactions in a gravityless, lower-dimensional spacetime. As a
famous example, the type IIB supergravity (or more generally the
type IIB superstring theory) on $AdS_5 \times {\bf S}^5$ is emergent
from the 4-dimensional $\CN =4$ supersymmetric $U(N)$ Yang-Mills
theory \ct{ads-cft}.\footnote{The overall $U(1)=U(N)/SU(N)$ factor
actually corresponds to the overall position of D3-branes and may be
ignored when considering dynamics on the branes, thereby leaving
only an $SU(N)$ gauge symmetry.} In our construction, $N \times N$
matrices are mapped to vector fields on some manifold $M$, so the
vector fields in Eq.\eq{local-vector} correspond to master fields of
large $N$ matrices \ct{master}, in other words, $(A_\mu, \Phi^a)
\sim N^2$. According to the AdS/CFT duality, we thus expect that the
metric \eq{emergent-metric} describes a deformed geometry induced by
excitations of the gauge and Higgs fields in the action
\eq{matrix-action}. For example, we may look for $1/2$ BPS
geometries in type IIB supergravity that arise from chiral primaries
of $\CN=4$ super Yang-Mills theory \ct{llm}. Recently this kind of
BPS geometries, the so-called bubbling geometry in AdS space, with a
particular isometry was completely determined in
\ct{llm}, where the $AdS_5 \times {\bf S}^5$ geometry
emerges from the simplest and most symmetric configuration. In next
section we will illustrate such kind of bubbling geometry described by
the metric \eq{emergent-metric} by considering self-dual configurations in the
gauge-Higgs system.

\section{Self-dual Einstein Gravity From Large $N$ Gauge Theory}

In the previous section we showed that the Ward's metric \eq{emergent-metric}
naturally emerges from the $D$-dimensional NC $U(1)$ gauge fields $A_M$
on $\IR^d_C \times \IR^{2n}_{NC}$ or equivalently the gauge-Higgs
system $(A_\mu,\Phi^a)$ in $d$-dimensional $U(N)$ gauge theory on $\IR^d_C$.
So, if an explicit solution for $A_M$ or $(A_\mu,\Phi^a)$ is known,
the corresponding metric \eq{emergent-metric} is, in principle, exactly
determined. However, it is extremely difficult to get a general solution
by solving the equations of motion for the action \eq{starting-action} or
\eq{matrix-action}. Instead we may try to solve a more simpler system
such as the first-order equations \eq{high-sd}, which are morally believed to
be `exactly solvable' in most cases. In this section we will further elucidate
the emergent gravity arising from gauge fields by showing that
the gauge-Higgs system $(A_\mu,\Phi^a)$ in half-BPS configurations
describes self-dual Einstein gravity. Since the case for $D=4$ and $n=2$ has
been extensively discussed in \ct{ys,hsy1,hsy2},
we will consider the other cases for $D \ge 4$.
For simplicity, the metrics in the action \eq{matrix-action} are supposed
to be the form already used in Eq.\eq{emergent-metric};
$g_{\mu\nu}=\delta_{\mu\nu}$ and $g_{ab}=\delta_{ab}$.

Note that the action \eq{starting-action} or \eq{matrix-action}
contains a background $B$, due to a uniform condensation of gauge
fields in a vacuum. But we will require a rapid fall-off of
fluctuating fields around the background at infinity in $\IR^D$ as
usual.\footnote{In the matrix representation
\eq{matrix-expansion}, this means that matrix components
$\Omega_{ij}(z)$ for the fluctuations are rapidly vanishing for $i,j
= N \to \infty$ as well as for $|z| \to \infty$, since roughly $N
\sim \vec{y} \cdot \vec{y}$.} Our boundary condition is $F_{MN} \to
0 $ at infinity. Eq.\eq{fncnc} then requires that $[x^a, x^b]_\star
\to i \theta^{ab}$ at $|y| \to \infty$. Thus the coordinates $y^a$
in \eq{cov-x} are vacuum expectation values of $x^a$ characterizing
the uniform condensation of gauge fields \ct{hsy2}. This
condensation of the $B$-fields endows the $\star$-algebra
$\CA_\theta$ with a remarkable property that translations act as an
inner automorphism of the algebra $\CA_\theta$ as shown in
Eq.\eq{inner-der}. But the gauge symmetry on NC spacetime requires
the covariant coordinates $x^a$ in Eq.\eq{cov-x} instead of $y^a$
\ct{madore}. The inner derivation ${\rm ad}_{D_a}$ in
Eq.\eq{adjoint-action} is then a `dual element' related to the
coordinate $x^a$. This is also true for the covariant derivatives
$D_\mu$ in \eq{cov-der} since they are related to $D_a= -i B_{ab}
x^b$ by the `matrix $T$-duality'; $D_a
\mapsto D_\mu$, as will be explained in Section 5.

It is very instructive to take an analogy with quantum mechanics.
Quantum mechanical time evolution in Heisenberg picture is defined
as an inner automorphism of the Weyl algebra obtained from a quantum
phase space
$$ f(t) = e^{iHt} f(0) e^{-iHt}$$ and
its evolution equation is of the form \eq{adjoint-action}
$$\frac{df(t)}{dt} = i[H, f(t)].$$
Here we liberally interpret $D_A(z,y)$ in Eq.\eq{adjoint-action} as
`multi-Hamiltonians' determining the {\it spacetime evolution} in
$\IR^D$. Then it is quite natural to interpret
Eq.\eq{adjoint-action} as a spacetime evolution equation determined
by the ``covariant Hamiltonians'' $D_A(z,y)$.

Let us be more precise about the meaning of the {\it spacetime
evolution}. If the Hamiltonian is slightly deformed, $H \to H +
\delta H$, the time evolution of a system is correspondingly
changed. Likewise, the fluctuation of gauge fields $A_M$ or $(A_\mu,
\Phi^a)$ around the background specified by $y^a$'s changes
$D_A(z,y)$, which in turn induces a deformation of the background
spacetime according to Eq.\eq{semi-vector}. This is precisely the
picture about the emergent geometry in \ct{hsy1,hsy2} and also a
dependable interpretation of the Ward's geometry
\eq{emergent-metric}. A consistent picture related to the AdS/CFT
duality was also observed in the last of Section 3.

For the above reason, all equations in the following will be
understood as inner derivations acting on $\CA_\theta$ like as
\eq{adjoint-action}. The adjoint action defined in this way
naturally removes a contribution from the background in the action
\eq{starting-action} or \eq{matrix-action} \ct{hsy1}.
For example, the first equation in \eq{hitchin} can be consistent
only in this way since the left hand side goes to zero at infinity
but the right hand side becomes $\sim \theta/\kappa^2$. It might be
remarked that this is the way to define the equations of motion in
the background independent formulation \ct{sw,seiberg} and thus it
should be equivalent to the usual NC prescription with
$\Phi_{MN}=0$.

\subsection{$D=4$ and $n=1$}

NC instanton solutions in this case were constructed in
\ct{chu-kly}. As was proved in Eq.\eq{general-sd}, NC $U(1)$ instantons are in
general equivalent to gravitational instantons. We thus expect that
the NC self-duality equations for $D=4$ and $n=1$ are mapped to
self-dual Einstein equations. We will show that the gauge-Higgs
system $(A_\mu,\Phi^a)$ in this case is mapped to two-dimensional
$U(\infty)$ chiral model, whose equations of motion are equivalent
to the Pleba\'nski form of the self-dual Einstein equations
\ct{q-park,ward,husain}.

We showed in Section 2 that 4-dimensional NC $U(1)$ gauge theory on
$\IR^2_C \times \IR^{2}_{NC}$ is mapped to 2-dimensional $U(N \to
\infty)$ gauge theory with the action \eq{matrix-action}. The
4-dimensional self-duality equations now become the $U(N \to
\infty)$ Hitchin equations on $\IR^2_C$:
\be \la{hitchin}
F_{\mu\nu} = \pm \half \vare_{\mu\nu} [\Phi, \Phi^\dagger],
\qquad D_\mu \Phi = \pm i \vare_{\mu\nu} D_\nu \Phi,
\ee
where $\Phi = \Phi_1 + i \Phi_2$. Note that the above equations also
arise as zero-energy solutions in $U(N)$ Chern-Simons gauge theory
coupled to a nonrelativistic complex scalar field in the adjoint
representation \ct{djpt}. It was shown in \ct{dunne} that the
self-dual system in Eq.\eq{hitchin} is completely solvable in terms
of Uhlenbeck's uniton method. A NC generalization of
Eq.\eq{hitchin}, the Hitchin's equations on $\IR^2_{NC}$, was also
considered in \ct{klee} with very parallel results to the
commutative case. We will briefly discuss the NC Hitchin's equations
in Section 5.

The equations \eq{hitchin} for the self-dual case (with $+$ sign)
can be elegantly combined into a zero-curvature condition
\ct{djpt,dunne} for the new connections defined by\footnote{Here we
will relax the reality condition of the fields $(A_\mu, \Phi^a)$ and
complexify them.}
\be \la{zero-connection}
\CA_{+} = A_+ + \Phi, \qquad
\CA_{-} = A_- - \Phi^\dagger
\ee
with $A_\pm = A_1 \pm iA_2$:
\be \la{zero-curvature}
\CF_{+-} = \p_+ \CA_- - \p_- \CA_+ - i [\CA_+ , \CA_-] = 0
\ee
where $\p_\pm = \p_1 \pm i\p_2$. Thus the new gauge fields should be
a pure gauge, that is, $\CA_\pm = i g^{-1} \p_\pm g$ for some $g \in
GL(N, \IC)$. Thus we can choose them to be zero, viz.
\be \la{gauge}
A_+ = - \Phi, \qquad A_- =  \Phi^\dagger.
\ee
Then the self-dual equations \eq{hitchin} reduce to
\bea \la{sde-1}
&& \p_+ \Phi^\dagger + \p_- \Phi + 2i [ \Phi, \Phi^\dagger ] = 0, \\
\la{sde-2}
&& \p_+ \Phi^\dagger - \p_- \Phi = 0.
\eea
Introducing another gauge fields $C_+ = - 2 \Phi$ and $C_- = 2
\Phi^\dagger$, Eq.\eq{sde-1} also becomes the zero-curvature
condition, hence $C_\pm$ are a pure gauge or
\be \la{zero-c}
\Phi = - \frac{i}{2} h^{-1} \p_+ h, \qquad \Phi^\dagger = \frac{i}{2}  h^{-1} \p_- h.
\ee
A group element $h(z)$ defines a map from $\IR^2_C$ to $GL(N, \IC)$
group, which is contractible to the map from $\IR^2_C$ to $U(N)
\subset GL(N, \IC)$. Then Eq.\eq{sde-2} implies that $h(z)$
satisfies the equation in the two-dimensional $U(N)$ chiral model
\ct{djpt,dunne}
\be \la{chiral-eq}
\p_+ (h^{-1} \p_- h) + \p_- (h^{-1} \p_+ h) = 0.
\ee

Eq.\eq{chiral-eq} is the equation of motion derived from the
two-dimensional $U(N)$ chiral model governed by the following
Euclidean action
\be \la{chiral-action}
S= \half \int d^2 z \Tr \p_\mu h^{-1} \p_\nu h \delta^{\mu\nu}.
\ee
A remarkable (mysterious) fact has been known
\ct{q-park,ward,husain} that in the $N \to \infty$ limit the chiral
model \eq{chiral-action} describes a self-dual spacetime whose
equation of motion takes the Pleba\'nski form of self-dual Einstein
equations \ct{plebanski}. Thus, including the case of $D=4$ and
$n=2$ in \ct{ys,hsy1,hsy2}, we have confirmed Eq.\eq{general-sd}
stating that the 4-dimensional self-dual system in the action
\eq{starting-action} or \eq{matrix-action} in general describes
the self-dual Einstein gravity where self-dual metrics are given by
Eq.\eq{emergent-metric}.

\subsection{$D=6$ and $n=1$}

Our current work has been particularly motivated by this case since
it was already shown in \ct{ym-instanton} that $SU(N)$ Yang-Mills
instantons in the $N \to \infty$ limit are gravitational instantons
too. Since NC $U(1)$ instantons are also gravitational instantons as
we showed before, it implies that there should be a close
relationship between $SU(N)$ Yang-Mills instantons and NC $U(1)$
instantons. A basic observation was the relation \eq{sun-sdiff},
which leads to the sound realization in Eq.\eq{matrix-action}. But
we will simply follow the argument in \ct{ym-instanton} for the
gauge group $G = U(N)$; in the meantime, we will confirm the results
for the emergent geometry from NC gauge fields.

Let us look at the instanton solution in $U(N)$ Yang-Mills theory.
The self-duality equation is given by
\be
F_{\mu\nu} = \pm \half \varepsilon_{\mu\nu\alpha\beta} F_{\alpha\beta}
\la{sdym}
\ee
where the field strength is defined by
\be
F_{\mu\nu} = \partial_\mu A_\nu - \partial_\nu A_\mu - i[A_\mu,A_\nu].
\ee
In terms of the complex coordinates and the complex gauge fields defined by
\bea
&& z_1 = \half(x^2 + i x^1), \qquad z_2 = \half(x^4 + i x^3), \xx
&& A_{z_1} = A_2 - i A_1, \qquad A_{z_2} = A_4 - i A_3, \nonumber
\eea
Eq.\eq{sdym} can be written as
\bea \la{holo}
&& F_{z_1 z_2}  = 0 = F_{\zbar_1 \zbar_2}, \\
\la{11-part}
&& F_{z_1 \zbar_1} \mp F_{z_2 \zbar_2 } = 0.
\eea

Now let us consider the anti-self-dual (ASD) case.  We first notice
that $F_{z_1 z_2} = 0$ implies that there exists a $u(N)$-valued
function $g$ such that $A_{z_a} = i g^{-1} \partial_{z_a} g \;(a=1,2)$.
Therefore one can choose a gauge
\be \la{gauge-fixing}
A_{z_a} = 0.
\ee
Under the gauge \eq{gauge-fixing}, the ASD equations lead to
\bea \la{asd1}
&& \partial_{\zbar_1} A_{\zbar_2} - \partial_{\zbar_2} A_{\zbar_1}
- i [A_{\zbar_1}, A_{\zbar_2}] = 0, \\
\la{asd2}
&& \partial_{z_1} A_{\zbar_1} + \partial_{z_2} A_{\zbar_2} = 0.
\eea
First notice a close similarity with Eqs.\eq{sde-1} and \eq{sde-2}.
Eq.\eq{asd2} can be solved by introducing a $u(N)$-valued function $\Phi$
such that
\be \la{sol-phi}
A_{\zbar_1} = - \partial_{z_2} \Phi, \qquad A_{\zbar_2} =
\partial_{z_1} \Phi.
\ee
Substituting \eq{sol-phi} into \eq{asd1} one finally gets
\be \la{asd-eq}
(\partial_{z_1} \partial_{\zbar_1}+
\partial_{z_2} \partial_{\zbar_2}) \Phi -
i [\partial_{z_1} \Phi, \partial_{z_2} \Phi] = 0.
\ee

Adopting the correspondence \eq{sun-sdiff}, we now regard $\Phi
\in u(N) \otimes C^\infty(\IR^4)$ in Eq.\eq{asd-eq} as a smooth function
on $\IR^4 \times \Sigma_g$, i.e., $\Phi = \Phi(x^\mu,p,q)$ where
$(p,q)$ are local coordinates of a two-dimensional Riemann surface
$\Sigma_g$. Moreover, a Lie algebra commutator is replaced by the
Poisson bracket \eq{poisson-bracket}
$$ \{f,g\} = \frac{\partial f}{\partial q}\frac{\partial g}{\partial p} -
 \frac{\partial f}{\partial p}\frac{\partial g}{\partial q},$$
that is,
\be \la{poisson}
[\Phi_1, \Phi_2] \to  i \{\Phi_1, \Phi_2 \},
\ee
where we absorbed $\theta$ into the coordinates $(p,q)$. After all,
the ASD Yang-Mills equation \eq{asd-eq} in the large $N$ limit is
equivalent to a single nonlinear equation in six dimensions
parameterized by $(x^\mu,p,q)$:
\be \la{6d-asd-eq}
(\partial_{z_1} \partial_{\zbar_1}+
\partial_{z_2} \partial_{\zbar_2}) \Phi +
\{\partial_{z_1} \Phi, \partial_{z_2} \Phi\} = 0.
\ee
Since Eq.\eq{6d-asd-eq} is similar to the well-known
second heavenly equation \ct{plebanski},
it was called in \ct{ym-instanton} as the six dimensional version of the
second heavenly equation.

Starting from $U(N)$ Yang-Mills instantons in four dimensions, we
arrived at the nonlinear differential equation for a single function
in six dimensions. It is important to notice that the resulting
six-dimensional theory is a NC field theory since the Riemann
surface $\Sigma_g$ carries a symplectic structure inherited from the
$u(N)$ Lie algebra through Eq.\eq{poisson} and it can be quantized
in general via deformation quantization \ct{kontsevich}. Since the
function $\Phi$ in \eq{6d-asd-eq} is a master field of $N \times N$
matrices \ct{master}, so $\Phi \sim N^2$, the AdS/CFT duality
\ct{ads-cft} implies that the master field $\Phi$ describes a
six-dimensional emergent geometry induced by Yang-Mills instantons.

To see the emergent geometry, consider an appropriate symmetry
reduction of Eq.\eq{6d-asd-eq} to show that it describes self-dual
gravity in four dimensions. There are many reductions from six to
four dimensional subspace leading to self-dual four-manifolds
\ct{ym-instanton}. A common feature is that the four dimensional
subspace necessarily contains the NC Riemann surface $\Sigma_g$. We
will show later how the symmetry reduction naturally arises from the
BPS condition in six dimensions. As a specific example, we assume
the following symmetry,
\be \la{symm-red}
\partial_{z_1} \Phi = \partial_{\zbar_1} \Phi,
\qquad \partial_{z_2} \Phi = \partial_{\zbar_2} \Phi,
\ee
or
\be \la{red-lambda}
\Phi(z_1,z_2,\zbar_1,\zbar_2,p,q) = \Lambda(z_1+\zbar_1 \equiv x,
z_2+\zbar_2 \equiv y,p,q).
\ee
Then Eq.\eq{6d-asd-eq} is precisely equal to the Husain's equation
\ct{husain} which is the reduction of self-dual Einstein equations
to the $sdiff(\Sigma_g)$ chiral field equations in two dimensions:
\be \la{husain}
\Lambda_{xx} + \Lambda_{yy} + \Lambda_{xq}\Lambda_{yp}-
\Lambda_{xp}\Lambda_{yq} = 0.
\ee
Note that we already encountered in Section 4.1 the two-dimensional
$sdiff(\Sigma_g)$ chiral field equations since $sdiff(\Sigma_g)
\cong u(N)$ according to the correspondence \eq{sun-sdiff}. We
showed in \ct{hsy2} that Eq.\eq{husain} can be transformed to the
first heavenly equation \ct{plebanski} which is a governing equation
of self-dual Einstein gravity. In the end we conclude that self-dual
$U(N)$ Yang-Mills theory in the large $N$ limit is equivalent to
self-dual Einstein gravity.

Now it is easy to see that the self-dual Einstein equation
\eq{husain} is coming from a 1/2 BPS equation in six dimensions
(see Eq.(34) in \ct{bkp}) defined by the first-order equation
\eq{high-sd}. According to our construction, the six-dimensional NC
$U(1)$ gauge theory
\eq{starting-action} is equivalent to the four-dimensional $U(N)$ gauge theory
\eq{matrix-action}. Therefore six-dimensional BPS equations can
be equivalently described by the gauge-Higgs system $(A_\mu,
\Phi^a)$ in the action \eq{matrix-action}. Let us newly denote the
NC coordinates $y^1, y^2$ and commutative ones $z^3, z^4$ as
$u^\alpha,\; \alpha=1,2,3,4$ while $z^1, z^2$ as $v^A,\; A=1,2$. The
1/2 BPS equations, Eq.(34), in \ct{bkp} can then be written as the
following form
\bea \la{six-bps-1}
&& F_{\a\b} = \pm \half \varepsilon_{\alpha\beta \gamma \delta}
F_{\gamma\delta}, \\
\la{six-bps-2}
&& F_{\a A} = F_{AB}=0.
\eea
Using Eqs.\eq{fcc}-\eq{fncnc}, the above equations can be rewritten
in terms of $(A_\mu, \Phi^a)$ where the constant term in \eq{fncnc}
can simply be dropped for the reason explained before.

$F_{AB}=0$ in Eq.\eq{six-bps-2} can be solved by $A_B = 0 \;
(B=1,2)$ and then $F_{\a A}=0$ demand that the gauge fields $A_\a$
should not depend on $v^A$. Thereby Eq.\eq{six-bps-1} precisely
reduces to the self-duality equation \eq{hitchin} for $D=4$ and
$n=1$. The symmetry reduction considered above is now understood as
the condition
\eq{six-bps-2}; in specific, the coordinates $v^A$ correspond to
$i(z_1- \zbar_1)$ and $i(z_2-\zbar_2)$ for the reduction
\eq{red-lambda}. However there are many different choices taking a
four-dimensional subspace in Eq.\eq{six-bps-1} which are related by
$SO(6)$ rotations \ct{bkp}. Unless $v^A \in (y^1,y^2)$, that is,
Eq.\eq{six-bps-1} becomes commutative Abelian equations in which
there is no non-singular solution, Eqs.\eq{six-bps-1} and
\eq{six-bps-2} reduce to four-dimensional self-dual Einstein equations, as
was shown in \ct{ym-instanton}. The above BPS equations also clarify
why the two-dimensional chiral equations in Section 4.1 reappear in
Eq.\eq{husain}.

\subsection{$D=8$ and $D=10$}

The analysis for the first-order system \eq{high-sd} becomes much
more complicated in higher dimensions. The unbroken supersymmetries
in $D=8$ have been analyzed in \ct{bkp}. Because the integrable
structure of Einstein equations in higher dimensions is little
known, it is difficult to precisely identify governing geometrical
structures emergent from the gauge theory \eq{starting-action} or
\eq{matrix-action} even for BPS states. Nevertheless some BPS
configurations can be easily implemented as follows. As we did in
Eqs.\eq{six-bps-1}-\eq{six-bps-2}, one can imbed the 4-dimensional
self-dual system for $n=1$ or $n=2$ into eight or ten dimensions.
The simplest case is that the metric \eq{emergent-metric} becomes
(locally) a product manifold $\CM_4 \times X$ where $\CM_4$ is a
self-dual (hyper-K\"ahler) four-manifold. For example, we can
consider an eight-dimensional configuration where
$(A_1,A_2,\Phi^3,\Phi^4)$ depend only on $(z^1,z^2, y^3, y^4)$
coordinates while $(\Phi^1,\Phi^2, A_3,A_4)$ do only on $(y^1,y^2,
z^3, z^4)$ in a $B$-field background with $\theta^{12} \neq 0$ and
$\theta^{34} \neq 0$, only non-vanishing components. There are many
similar configurations. We will not exhaust them, instead we will
consider the simplest cases which already have some relevance to
other works.

The simplest BPS state in $D=8$ is the case with $n=2$ in the action
\eq{matrix-action}; see Eq.(55) in \ct{bkp}. The equations are of the form
\bea \la{8-bps-1}
&& F_{\mu\nu} = \pm \half \varepsilon_{\mu\nu\lambda\sigma}
F_{\lambda\sigma}, \\
\la{8-bps-2}
&& [\Phi^a,\Phi^b] = \pm \half \varepsilon_{abcd}  [\Phi^c,\Phi^d], \\
\la{8-bps-3}
&& D_\mu \Phi^a = 0.
\eea
A solution of Eq.\eq{8-bps-3} is given by $A_\mu = A_\mu(z)$ and
$\Phi^a = \Phi^a(y)$. Then Eq.\eq{8-bps-1} becomes commutative Abelian
equations which allow no non-singular solutions, while
\eq{8-bps-2} reduces to Eq.\eq{sd-higgs} describing 4-dimensional self-dual
manifolds \ct{hsy1}. Thus the metric \eq{emergent-metric} in this
case leads to a half-BPS geometry $\IR^4 \times \CM_4$. Since we
don't need instanton solutions in Eq.\eq{8-bps-1}, we may freely
replace $\IR^4$ by 4-dimensional Minkowski space $\IR^{1,3}$ (see
the footnote \ref{metric-convention}).

The above system was considered in \ct{d3-d7} in the context of
D3-D7 brane inflationary model. The model consists of a D3-brane
parallel to a D7-brane at some distance in the presence of $\CF_{ab}
= (B + F)_{ab}$ on the worldvolume of the D7-brane, but transverse
to the D3-brane. The $\CF$-field plays the role of the
Fayet-Illiopoulos term from the viewpoint of the D3-brane
worldvolume field theory. Because of spontaneously broken
supersymmetry in de Sitter valley the D3-brane is attracted towards
the D7-brane and eventually it is dissolved into the D7-brane as a
NC instanton. The system ends in a supersymmetric Higgs phase with a
smooth instanton moduli space. An interesting point in \ct{d3-d7} is
that there is a relation between cosmological constant in spacetime
and noncommutativity in internal space. Our above result adds a
geometrical picture that the internal space after tachyon
condensation is developed to a gravitational instanton, e.g., an ALE
space or $K3$.

Another interesting point, not mentioned in \ct{d3-d7}, is that it effectively
realizes the dynamical compactification of extra dimensions suggested in
\ct{dy-com}. Since the D3-brane is an instanton inside the D7-brane,
particles living in the D3-brane are trapped in the core of the
instanton with size $\sim \theta^2$ where the noncommutativity scale
$\theta$ is believed to be roughly Planck scale. Since the instanton
(D3-brane) results in a spontaneous breaking of translation symmetry
and supersymmetry partially, Goldstone excitations corresponding to
the broken bosonic and fermionic generators are zero-modes trapped
in the core of the instanton. ``Quarks" and ``leptons" might be
identified with these fermionic zero-modes \ct{dy-com}.

We argued in the last of Section 3 that the 10-dimensional metric
\eq{emergent-metric} for $d=4$ and $n=3$ reasonably corresponds to
an emergent geometry arising from the 4-dimensional $\CN = 4$
supersymmetric $U(N)$ Yang-Mills theory. Especially it may be
closely related to the bubbling geometry in AdS space found by Lin,
Lunin and Maldacena (LLM) \ct{llm}. One may notice that the LLM
geometry is a bubbling geometry deformed from the $AdS_5 \times {\bf
S}^5$ background which can be regarded as a vacuum manifold emerging
from the self-dual RR five-form background, while the Ward's
geometry \eq{emergent-metric} is defined in a 2-form $B$-field
background and becomes (conformally) flat if all fluctuations are
turned off, say, $(A_\mu, \Phi^a) \to (0, y^a/\kappa)$. But it turns
out that the LLM geometry is a special case of the Ward's geometry
\eq{emergent-metric}.

To see this, recall that the $AdS_5 \times {\bf S}^5$ background
is conformally flat, i.e.,
\be \la{ads5-s5}
ds^2 = \frac{L^2}{\rho^2}(\eta_{\mu\nu} dz^\mu dz^\nu + dy^a dy^a) =
\frac{L^2}{\rho^2}(\eta_{\mu\nu} dz^\mu dz^\nu + d\rho^2) + L^2 d \Omega_5^2
\ee
where $\rho^2 = \sum_{a=1}^6 y^a y^a$ and $d \Omega_5^2$ is the
spherically symmetric metric on ${\bf S}^5$. It is then easy to see
that the metric \eq{ads5-s5} is exactly the vacuum geometry of
Eq.\eq{emergent-metric} when the volume form $\omega$ in
Eq.\eq{v-form} is given by
\be  \la{ads-vol}
\omega = \frac{dy^1 \wedge \cdots \wedge dy^6}{\rho^2}.
\ee
Therefore it is obvious that the Ward's metric \eq{emergent-metric}
with the volume form \eq{ads-vol} describes a bubbling geometry
which approaches to the $AdS_5 \times {\bf S}^5$ space at infinity
where fluctuations are vanishing, namely, $(A_\mu, \Phi^a) \to (0,
y^a/\kappa)$. Note that the flat spacetime $\IR^{1,9}$ is coming
from the volume form $\omega = dy^1 \wedge \cdots \wedge dy^6$, so
Eq.\eq{ads-vol} should correspond to some nontrivial soliton
background from the gauge theory point of view. We will discuss in
Section 5 a possible origin of the volume form \eq{ads-vol}.

Now let us briefly summarize half-BPS geometries of type IIB string
theory corresponding to the chiral primaries of ${\cal N} = 4$ super
Yang-Mills theory \ct{llm}. These BPS states are giant graviton
branes which wrap an ${\bf S}^3$ in $AdS_5$ or an ${\widetilde {\bf
S}}^3$ in ${\bf S}^5$. Thus the geometry induced (or back-reacted)
by the giant gravitons preserves $SO(4) \times SO(4) \times R$
isometry. It turns out that the solution is completely determined by
a single function which is specified with two types of boundary
conditions on a particular plane corresponding to either of two
different spheres shrinking on the plane in a smooth fashion. The
LLM solutions are thus in one-to-one correspondence with various
2-colorings of a 2-plane, usually referred to as `droplets' and the
geometry depends on the shape of the droplets. The droplet
describing gravity solutions turns out to be the same droplet in the
phase space describing free fermions for the half-BPS states.

The solutions can be analytically continued to those with $SO(2,2)
\times SO(4) \times U(1)$ symmetry \ct{llm}, so the solutions have
the $AdS_3 \times {\bf S}^3$ factor rather than ${\bf S}^3 \times
{\widetilde {\bf S}}^3$. After an analytic continuation, a
underlying 4-dimensional geometry $\CM_4$ attains a nice geometrical
structure at asymptotic region, where $AdS_3 \times {\bf S}^3 \to
\IR^{1,5}$ and $\CM_4$ reduces to a hyper-K\"ahler geometry. But
it loses the nice picture in terms of fermion droplet since the
solution is now specified by one type of boundary condition. It is
interesting to notice that the asymptotic bubbling geometry for the
type IIB case is the Gibbons-Hawking metric \ct{gibb-hawk} and the
real heaven metric \ct{real-heaven} for the M theory case, which are
all solutions of NC electromagnetism \ct{hsy1,hsy2}.

It is quite demanding to completely determine general half-BPS
geometries emerging from the gauge-Higgs system in the action
\eq{matrix-action}. Hence we will look at only an asymptotic
geometry (or a local geometry) which is relatively easy to identify.
For the purpose, we consider the $n=3$ case on 4-dimensional
Minkowski space $\IR^{1,3}$. It is simple to mimic the previous
half-BPS configurations in $D=6, 8$ with trivial extra Higgs fields.
Then the resulting metric \eq{emergent-metric} will be locally of
the form $\CM_4  \times \IR^{1,5}$ akin to the asymptotic bubbling
geometry. However $\CM_4$ can be a general hyper-K\"ahler manifold.
Therefore the solutions we get will be more general, whose explicit
form will depend on underlying Killing symmetries and boundary
conditions. For example, the type IIB case is given by a
hyper-K\"ahler geometry with one translational Killing vector
(Gibbons-Hawking) while the M theory case is with one rotational
Killing vector (real heaven) \ct{bakas}. Therefore we may get in
general bubbling geometries in the M theory as well as the type IIB
string theory.

\section{Discussion}

We showed reasonable evidences that the 10-dimensional metric
\eq{emergent-metric} for $d=4$ and $n=3$ describes the emergent
geometry arising from the 4-dimensional $\CN = 4$ supersymmetric
$U(N)$ Yang-Mills theory and thus might explain the AdS/CFT duality
\ct{ads-cft}. An important point in this context is that the volume
form \eq{ads-vol} is required to describe the $AdS_5 \times {\bf
S}^5$ background. What is the origin of this nontrivial volume form
? In other words, how to realize the self-dual RR five-form
background from the gauge theory point of view ?

To get some hint about the question, first note that the $AdS_5
\times {\bf S}^5$ geometry emerges from multi-instanton collective
coordinates which dominates the path integral in a large $N$ limit
\ct{green-dorey}. The factor $d^4 z d\rho \rho^{-5}$ appears in
bosonic collective coordinate integration (with $z^\mu$ the
instanton 4-positions) which agrees with the volume form of the
conformally invariant space $AdS_5$, where instanton size
corresponds to the radial coordinate $\rho$ in Eq.\eq{ads5-s5}.
Another point is that the $AdS_5 \times {\bf S}^5$ space corresponds
to the LLM geometry for the simplest and most symmetric
configuration which reduces to the usual Gibbons-Hakwing metric
\eq{gibbons-hawking} at asymptotic regions \ct{llm}. This result is
consistent with the picture in Section 4.2 that $U(N)$ instantons at
large $N$ limit are indeed gravitational instantons. It is then
tempted to speculate that the $AdS_5 \times {\bf S}^5$ geometry
would be emerging from a maximally supersymmetric instanton solution
of Eq.\eq{high-sd} in $D=10$. It should be an interesting future
work.

In addition, we would like to point out that an $AdS_p \times {\bf S}^q$
background arises from Eq.\eq{emergent-metric} in the same way as
Eq.\eq{ads-vol} by choosing the volume form $\omega$ as follows
\be  \la{ads-p-vol}
\omega = \frac{dy^1 \wedge \cdots \wedge dy^{q+1}}{\rho^2}
\ee
with $\rho^2 = \sum_{a=1}^{q+1} y^a y^a$ and $(A_\mu, \Phi^a) = (0, y^a/\kappa)$.
A particularly interesting case is $d=2$ and $n=2$ for which
the volume form \eq{ads-p-vol} leads to the $AdS_3 \times {\bf S}^3$ background and
the action \eq{matrix-action} describes matrix strings \ct{matrix-string,m-review}.
We believe that the metric \eq{emergent-metric} with $\omega = dy^1 \wedge \cdots
\wedge dy^{4}/\rho^2$ describes a bubbling geometry emerging from the matrix strings.

One might already notice a subtle difference between the matrix
action \eq{matrix-action} and the Ward's metric \eq{emergent-metric}.
According to our construction in Section 2, the number of the Higgs
fields $\Phi^a$ is even while the Ward construction has no such
restriction. But it was shown in \ct{hsy1,hsy2} that the Gibbons-Hawking metric
\eq{gibbons-hawking} for the $d=3$ and $m=1$ case also arises from
the $d=0$ and $m=4$. It implies that we can replace some transverse
scalars by gauge fields and vice versa. Recalling that the fields in
the action \eq{matrix-action} are all $N \times N$ matrices, of
course, $N \to \infty$, it is precisely `matrix $T$-duality'
exchanging transverse scalars and gauge fields associated with a
compact direction in $p$-brane and $(p+1)$-brane worldvolume
theories through (see Eq.(154) in \ct{m-review})
\be \la{t-dual}
\Phi^a  \leftrightarrow i D_\mu = i (\p_\mu -i A_\mu).
\ee
With this identification, the $d$-dimensional $U(N)$ gauge theory
\eq{matrix-action} can be obtained by applying the $d$-fold `matrix $T$-duality' \eq{t-dual}
to the $0$-dimensional IKKT matrix model \ct{ikkt,japan}
\be \la{ikkt-action}
S = - \frac{2\pi \kappa^2}{g_s} \Tr \left(
\frac{1}{4}g_{MP}g_{NQ}[\Phi^M, \Phi^N][\Phi^P, \Phi^Q] \right).
\ee

However, the $T$-duality \eq{t-dual}  gives rise to qualitatively
radical changes in worldvolume theory. First it changes the
dimensionality of the theory and thus it affects its
renormalizability (see Sec. VI in \ct{m-review} and references
therein for this issue in Matrix theory). For example, the action
\eq{matrix-action} for $d > 4$ is not renormalizable since the coupling
constant $g^2_{YM} \sim g_s \kappa^{\frac{d-4}{2}} \sim g_s
m_s^{4-d}$ has negative mass dimension in this case. Second it also
changes a behavior of the emergent metric \eq{emergent-metric}. But
these changes are rather consistent with the fact that under the
$T$-duality \eq{t-dual} a D$p$-brane is transformed into a
D$(p+1)$-brane and vice versa.

Our construction in Section 2 raises a bizarre question about the
renormalization property of NC field theory. If we look at the action
\eq{starting-action}, the theory superficially seems to be non-renormalizable for $D>4$
since the coupling constant \eq{coupling1} has negative mass
dimension. But this non-renormalizability appears as a fake if we
use the matrix representation \eq{matrix-expansion} together with
the redefinition of variables in Eq.\eq{open}. The resulting
coupling constant, denoted as $g_d$, in the final action
\eq{matrix-action} depends only on the dimension of commutative
spacetime rather than the entire spacetime. Since the resulting
$U(N)$ theory is in the limit $N \to \infty$, while the 't Hooft
coupling $\lambda \equiv g^2_d N$ is kept fixed, planar diagrams
dominate in this limit \ct{1/N}. Since the dependence of NC
coordinates in the action \eq{starting-action} has been encoded into
the matrix degrees of freedom, one may suspect that the divergence
of the original theory might appear as a divergence of perturbation
series as a whole in the action \eq{matrix-action}. The convergence
aspect of the planar perturbation theory concerns $N_p(n)$, the
number of planar diagrams in $n$th order in $\lambda$. It was shown
in \ct{planar} that $N_p(n)$ behaves asymptotically as
\be \la{asymp-diagram}
N_p(n) \stackrel{n \to \infty} {\sim} c^n, \qquad c={\rm constant}.
\ee
Therefore the planar theory (unlike the full theory) for $d \le 4$ has a
formally convergent perturbation series, provided the ultraviolet and
infrared divergences of individual diagrams are cut off \ct{planar}.
It will be interesting to carefully examine the renormalization property
of NC field theories along this line.

We showed in Section 3 that the Ward metric \eq{emergent-metric} is
emerging from commutative, i.e., $\CO(\theta)$, limit. Since the
vector fields in Eq.\eq{semi-vector} are in general higher order
differential operators acting on $\CA_\theta$, we thus expect that
they actually define a `generalized gravity' beyond Einstein
gravity, e.g., the NC gravity \ct{nc-gravity} or the NC unimodular
gravity \ct{nc-um-gravity}.\footnote{The latter seems to be quite
relevant to our emergent gravity since the vector fields $V_M$ in
Eq.\eq{semi-vector} always belong to the volume preserving
diffeomorphisms, which is a generic property of vector fields
defined in NC spacetime. It should be interesting to more clarify
the relation between the NC unimodular gravity \ct{nc-um-gravity}
and the emergent gravity.} It was shown in \ct{hsy2} that the
leading derivative corrections in NC gauge theory start with four
derivatives, which was conjectured to give rise to higher order
gravity. As was explicitly checked for the self-dual case, Einstein
gravity maybe emerges from NC gauge fields in commutative limit,
which then implies that the leading derivative corrections give rise
to higher order terms with four more derivatives compared to the
Einstein gravity. This means that the higher order gravity starts
from the second order corrections in $\theta$ with higher
derivatives, that is, no first order correction in $\theta$ to the
Einstein gravity. Interestingly this result is consistent with those
in \ct{nc-gravity} and also in \ct{nc-correction} calculated from
the context of NC gravity.

It was shown in Section 4.1 that the self-duality system for the $D=4$ and
$n=1$ case is mapped to the two-dimensional $U(\infty)$ chiral model
\eq{chiral-action} which is remarkably equivalent to self-dual
Einstein gravity \ct{q-park,ward,husain}. But this case should not
be much different from the $D=4$ and $n=2$ case in \ct{hsy1,hsy2}
since they equally describe the self-dual Einstein gravity. Indeed
we can make them bear a close resemblance each other. For the
purpose, let us consider a four-dimensional NC space $\IR^2_{NC}
\times \IR^2_{NC}$. We can choose the matrix representation
\eq{matrix-expansion} only for the second factor, i.e.,
\be \la{partial-matrix-expansion}
\phi(y^1,y^2,y^3,y^4) = \sum_{i,j} \Omega_{ij} \; (y^1,y^2)
|i \rangle \langle j |.
\ee
As a result, the action \eq{matrix-action} now becomes
two-dimensional $U(N)$ gauge theory on $\IR^2_{NC}$. The self-dual
equations in Eq.\eq{hitchin} in this case are given by the NC
Hitchin equations, now defined on $\IR^2_{NC}$ instead of
$\IR^2_{C}$. The NC Hitchin equations have been considered by K. Lee
in \ct{klee} with very parallel results with the commutative case
\eq{hitchin}. It is interesting that there exist two different
realizations for self-dual Einstein gravity, whose relationship
should be more closely understood.

Finally it will be interesting to consider a compact NC space
instead of $\IR^{2n}_{NC}$, for instance, a NC $2n$-torus ${\bf
T}^{2n}_{NC}$. Since the module over a NC torus is still infinite
dimensional \ct{sw}, the matrix representation \eq{matrix-expansion}
is also infinite dimensional. Thus we expect that our construction
in Section 2 and 3 can be applied even to the NC torus without many
essential changes.

\section*{Acknowledgments}

After posting this paper to the arXiv, we were informed of related
works on YM-Higgs BPS configurations on NC spaces \ct{lechtenfeld}
and on the relation between a large $N$ gauge theory, a Moyal
deformation and a self-dual gravity \ct{castro} by O. Lechtenfeld
and C. Castro, respectively. We thank them for the references. This
work was supported by the Alexander von Humboldt Foundation.

\newpage

%%%%%%%%%%%%%%%%% Journal Macros %%%%%%%%%%%%%%%%%%%%%%%%%%%

\nc{\npb}[3]{Nucl. Phys. {\bf B#1}, #2 (#3)}

\nc{\plb}[3]{Phys. Lett. {\bf B#1}, #2 (#3)}

\nc{\prl}[3]{Phys. Rev. Lett. {\bf #1}, #2 (#3)}

\nc{\prd}[3]{Phys. Rev. {\bf D#1}, #2 (#3)}

\nc{\ap}[3]{Ann. Phys. {\bf #1}, #2 (#3)}

\nc{\prep}[3]{Phys. Rep. {\bf #1}, #2 (#3)}

\nc{\epj}[3]{Eur. Phys. J. {\bf #1}, #2 (#3)}

\nc{\ptp}[3]{Prog. Theor. Phys. {\bf #1}, #2 (#3)}

\nc{\rmp}[3]{Rev. Mod. Phys. {\bf #1}, #2 (#3)}

\nc{\cmp}[3]{Commun. Math. Phys. {\bf #1}, #2 (#3)}

\nc{\mpl}[3]{Mod. Phys. Lett. {\bf #1}, #2 (#3)}

\nc{\cqg}[3]{Class. Quant. Grav. {\bf #1}, #2 (#3)}

\nc{\jhep}[3]{J. High Energy Phys. {\bf #1}, #2 (#3)}

\nc{\atmp}[3]{Adv. Theor. Math. Phys. {\bf #1}, #2 (#3)}

\nc{\hepth}[1]{{\tt hep-th/{#1}}}

%%%%%%%%%%%%%%%%%%%%%%%%%%%%%%%%%%%%%%%%%%%%%%%%%%%%%%%%%%%

\end{document}